\documentstyle[floats,prd,aps]{revtex}
\begin{document}
\draft

\title{Collisions of boosted black holes: perturbation
theory prediction of gravitational radiation}

\author{Andrew M.\ Abrahams and Gregory B.\ Cook}
\address{Center for Radiophysics and Space Research,
         Cornell University, Ithaca, NY 14853}

\date{\today}

\twocolumn[
\maketitle

\begin{abstract}
\widetext
We consider general relativistic Cauchy data representing
two nonspinning, equal-mass black holes boosted toward each other.
When the black holes are close enough to each other and
their momentum is sufficiently high, an
encompassing apparent horizon is present so the system can be
viewed as a single, perturbed black hole. We employ gauge-invariant
perturbation theory, and integrate the Zerilli equation
to analyze these time-asymmetric data sets and compute
gravitational wave forms and emitted energies.
When coupled with a simple Newtonian analysis of the infall
trajectory, we find striking agreement between the
perturbation calculation of emitted
energies and the results of fully general relativistic
numerical simulations of time-symmetric initial data.
\end{abstract}
]

\narrowtext
\section{Introduction}
\label{sec:intro}
The collision of two black holes is expected to
be an important source of gravitational radiation
for gravity wave detectors currently under
construction.  A major theoretical effort is underway to compute the
gravitational wave form from the orbiting inspiral
and coalescence of black-hole binaries.
Post-Newtonian theory should provide a sufficiently
accurate wave form for much of the inspiral phase and enable
the extraction of considerable information about the parameters
of the system\cite{cutler_etal93}.
However, it is anticipated that a fully general relativistic
treatment will be necessary to predict the wave form
from the final stages of coalescence.

Recently, the seminal calculations of Smarr and
Eppley\cite{smarr77} for the head-on collision of
two equal-mass holes starting at rest from a finite separation
have been redone\cite{anninos_etal93} with the benefit
of new numerical techniques and theoretical
tools, as well as vastly increased
computational resources.   The basic
conclusions reached by the modern calculations are remarkable
for their similarity to those of the original study.
The maximum amount of energy radiated is small, less than
0.1\% of the mass of the system.  Also, the wave forms are
indistinguishable at the level of the numerical accuracy
from quasinormal mode oscillations of a black hole.

Compelled by these results, Price and Pullin\cite{price_pullin94}
analyzed the Misner initial data for two
black holes at a moment of time-symmetry  as if it
represented a single, perturbed black hole.  Using gauge-invariant
perturbation theory and the Zerilli equation,
they were able to compute the initial distortion of the
black hole, and the resulting asymptotic wave forms and
energy fluxes as a function of separation.
For small separation, when the approximation of
the merged system as a single, perturbed black hole is
expected to be most valid, the agreement between
the radiated energy from
perturbation theory and the results of fully relativistic
simulations\cite{price_pullin94} is
excellent, apparently within the error bars of the
numerical calculation.  The quadrupole wave forms are
also remarkably similar when read off at
the same radius. Only when the initial separation
of the black holes is somewhat larger than the cutoff
for encompassing apparent horizons is there
substantial discrepancy
between the perturbation theory and evolution results
for radiation efficiency.

Previously (Ref.~\cite{cookabrahams92}, hereafter Paper 1),
the current authors had used a similar perturbation
theory analysis to study spurious radiation in time-symmetric
and asymmetric two-black-hole data sets in the opposite
limit -- that of large separation.   These data
sets represent the direct extension of Misner data allowing
the two black holes to have nonvanishing {\em initial}\,
linear and angular momenta.  In this paper, we apply these
gauge-invariant perturbation techniques, in the spirit of the
Price and Pullin paper, to the close limit when the two
black holes have an encompassing apparent horizon, and examine
the gravitational radiation wave form and the amount of energy
radiated for time-asymmetric initial data.

There are several motivations for this study
of boosted black-hole initial data.
In the final plunge phase of binary black hole
coalescence, the black holes are likely to have
substantial infall velocities.  Although the actual collision
is not expected to be head-on, axisymmetric calculations
of boosted head-on collisions are an interesting
limiting case of full three-dimensional plunge simulations.
In addition, these calculations extend the regime where
perturbation calculations can be fairly compared with the
fully relativistic simulations.  It is clear that, for
time-symmetric initial data with no event horizon, the
perturbation theory calculation should greatly overestimate the
distortion of the merged black hole, and thus the
radiated energy.  For these cases the merged black hole
doesn't form until the individual black holes have
evolved toward each other and have some infall momentum.
It is this merged black hole that one would like to
analyze with perturbation theory and compare against
the evolution results.

\section{Methodology}
\label{sec:methods}
In this paper, we will be concerned with two equal-mass
nonrotating black holes, each with axisymmetric inward-pointing
momentum $P$ (the slice has zero net-momentum).
Initial-data sets representing one or more black holes
with individually specifiable linear and angular momenta
are constructed using the conformal-imaging approach
developed by York and coworkers\cite{conf_imag}.
The case of two black holes with axisymmetric momenta was
implemented numerically by Cook\cite{cook91}.  The data sets used
for this study were constructed using a code based on the
\v{C}ade\v{z}-coordinate approach described in that work.
The reader should refer there for details regarding the
construction of the data sets and for further descriptions
of the parameterization of the data sets described below.
The separation of the holes is parameterized by $\beta$,
which is related to the
bispherical-coordinate separation parameter $\mu_0$
by the relation $\mu_0 = \cosh^{-1} (\beta/2)$ in the
case of equal-mass holes.
The code is used to compute the inversion-symmetric
(with minus isometry condition) extrinsic curvature $K_{ij}$
and to solve the Hamiltonian constraint for the conformal
factor $\psi$.  Once the full initial data is computed,
several physical quantities characterizing the system are
computed.  Of interest in this paper are the ADM mass of the
initial-data slice $M$, the proper separation of the holes $\ell$,
and the masses of the individual holes $m_1 = m_2$ defined in
terms of the area of the marginally outer-trapped surface
associated with each hole.  We also define the total or bare
mass of the system by $m = m_1 + m_2$.  Note that the
difference between $m$ and $M$ is due to the binding
energy of the system.
Given an initial-data set, we use the boundary-value-problem
method described in Paper 1 to locate all marginally
outer-trapped surfaces surrounding the two holes (if they exist)
and identify the apparent horizon(s).
We should note that there is nothing unique
about our choice of initial data for representing
two colliding black holes.  One could imagine, for example,
considering data with Euclidean topology with the black holes
represented by boosted matter collapsed inside its
horizon.  Our initial data was chosen for the convenience of
its highly refined numerical treatment\cite{cook91} and
earlier physical exploration (Paper 1).

Like Price and Pullin\cite{price_pullin94},
for purposes of our analysis we treat the spacetime as
a perturbation (but not necessarily a time-symmetric one) of
Schwarzschild. First, we establish a Schwarzschild-like
coordinate system around the two black holes in terms of
the (background space) isotropic coordinates used in the
numerical solution $r=r_i(1+M/2r_i)^2$.  Note that the background
space of the numerical solution can be directly parameterized by
the isotropic radial coordinate $r_i$ even thought the numerical
solution is found in \v{C}ade\v{z} coordinates.
The total ADM mass of the slice $M$ is used as the Schwarzschild
background mass.  Tortoise coordinates are also defined in the
usual way: $r_* = r + 2M \ln (r/2M -1)$.  Computation of wave
perturbations involves the calculation of multipole amplitudes
by surface integrals.  These are performed over constant Schwarzschild
radial-coordinate two-spheres.  The integrands involve
the conformal factor $\psi$, Schwarzschild-coordinate
extrinsic-curvature components $K_{ij}$, and their
Schwarzschild-coordinate radial derivatives.  Calculation of these
quantities at their required locations is achieved with bi-cubic
spline interpolations and a series of coordinate transformations.

The gauge-invariant function $Q_{\ell m}$ is formed out
of multipole projections of $\psi$ and $\psi_{,r}$ computed by
numerical integrations over a coordinate
two-sphere~(cf. Refs.~\cite{moncrief74,cpm79,ancsa92}).
For this paper we compute only the case of $Q\equiv Q_{20}$.
We also require the Schwarzschild time-derivative of the
gauge-invariant function $\partial_t Q$.
This time derivative is computed as
\begin{equation}
\partial_t Q = \alpha {\cal L}_{\bf n} Q,
\end{equation}
where ${\cal L}_{\bf n}$ is the Lie derivative along
the slice-normal congruence ${\bf n}$ and the
factor $\alpha=\sqrt{(1-2M/r)}$
comes about from the transformation from the slice-normal time
coordinate to the Schwarzschild time-coordinate.
The Lie derivative of $Q$ is calculated using the
extrinsic curvature (and its radial derivative) via the definition
\begin{equation}
{\cal L}_{\bf n} g_{ij} = -2 K_{ij}.
\end{equation}

The gauge-invariant perturbation function and
its time derivative, known as a function of radius
surrounding the merged black hole, serves as
initial data for integration of the Zerilli equation.
The numerically generated initial perturbation is
interpolated onto a fine grid (typically 8000-16000 zones)
that is even in $r_*$ and extends from $r_*=-500M$ to
$r_*=2000M$.  The Zerilli
equation (cf. Ref.~\cite{moncrief74,ancsa92,price_pullin94})
is then integrated forward in time until the whole perturbation
has been propagated to $|r_*| \rightarrow  \infty$.
Approximate asymptotic wave forms and energy fluxes are computed
at large radii.

Our code for calculating the initial black-hole perturbation
from numerically generated initial data was checked
by comparing it against the time-symmetric results
of Price and Pullin\cite{price_pullin94}.
It should be noted that they analytically expanded the metric
perturbation about Schwarzschild in powers of the
parameter $\epsilon = 1/|\ln \mu_0 |$ and retain only the
leading term in $\epsilon$.  In the limit of small separation, the
initial perturbation we obtain numerically agrees closely with
their analytic results (for $\beta\lesssim3.25$ the agreement is
better than $5\%$).
Not surprisingly, for larger separations the differences
become larger.  For the horizon cutoff point of
$\beta=4.17$ or $\mu_0=1.36$ (the largest separation allowing an
encompassing apparent horizon for time-symmetric initial data),
the analytic prediction
for the energy radiated is about $60\%$ higher than
the result from the full solution.
It is interesting to note that neglecting the higher-order
terms in $\epsilon$ always seems to lead to a
greater amount of radiated energy.

\section{Waveforms and energy flux}
\label{seq:wave_energy}
The perturbed black-hole approximation assumed in this paper
is not valid if the two black holes have not merged (have no
common event horizon).  For separations
small enough that an encompassing apparent horizon exists, we
find that the addition of inward-pointing linear momenta makes
the encompassing apparent horizon more spherical and the
maximum radiation efficiency (defined as the ratio of M minus
the mass of the apparent horizon to M) decreases.  Moreover, we
find that the metric perturbation $Q$ always gets smaller.  We,
therefore,
contend that our treatment of time-asymmetric initial-data sets
with inward-pointing linear momenta is always at least as valid
as the study of the time-symmetric solution.

\begin{figure}
\special{hscale=0.45 vscale = 0.45 hoffset = -0.2 voffset = -4.25
psfile=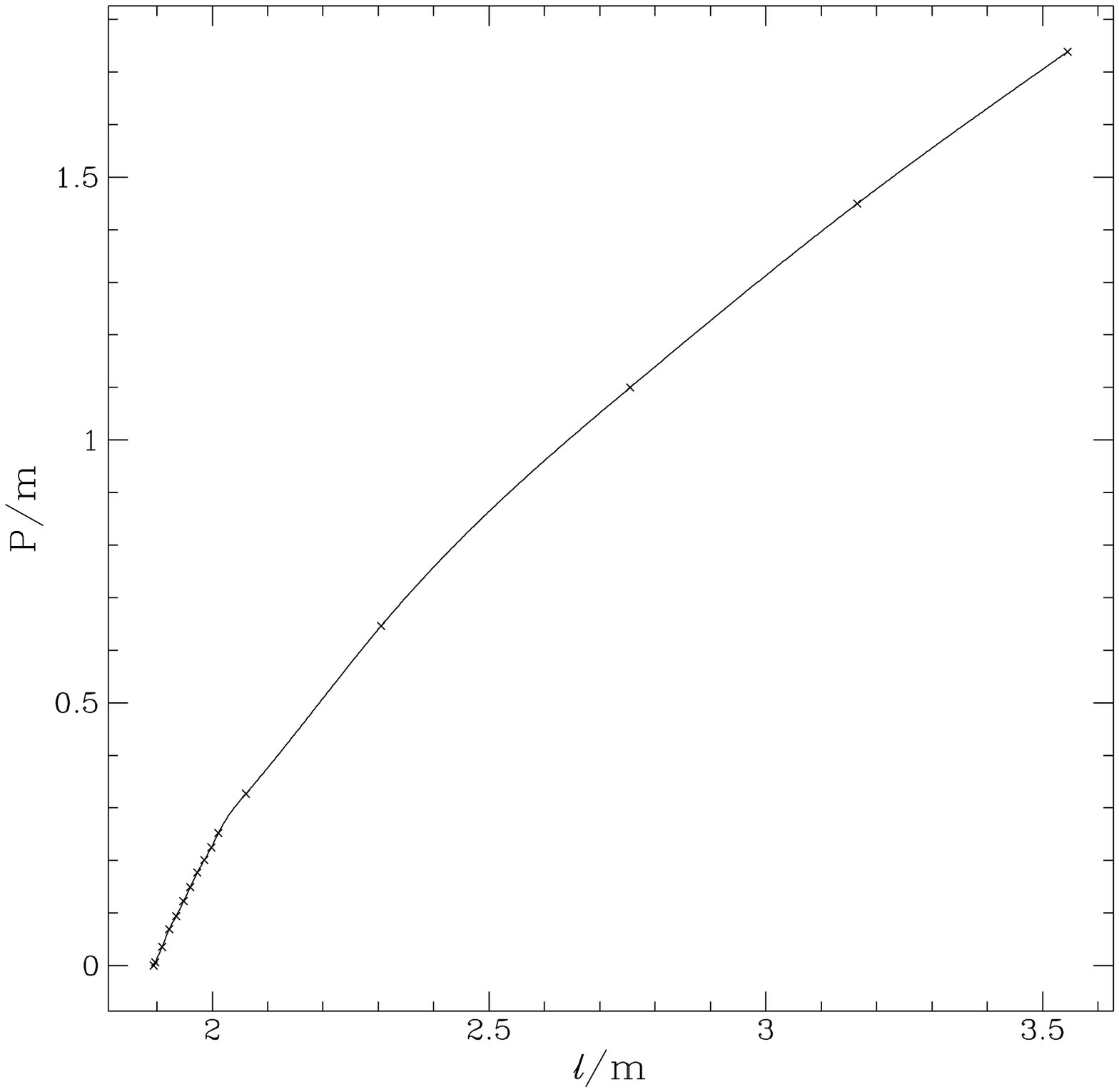}
\vspace{3.4in}
\caption{The apparent horizon
formation line.  The inward linear momentum
on each hole $P/m$ is plotted as a function
of proper separation $\ell/m$.}
\label{fig:hl}
\end{figure}

\begin{figure}
\special{hscale=0.45 vscale = 0.45 hoffset = -0.2 voffset = -4.25
psfile=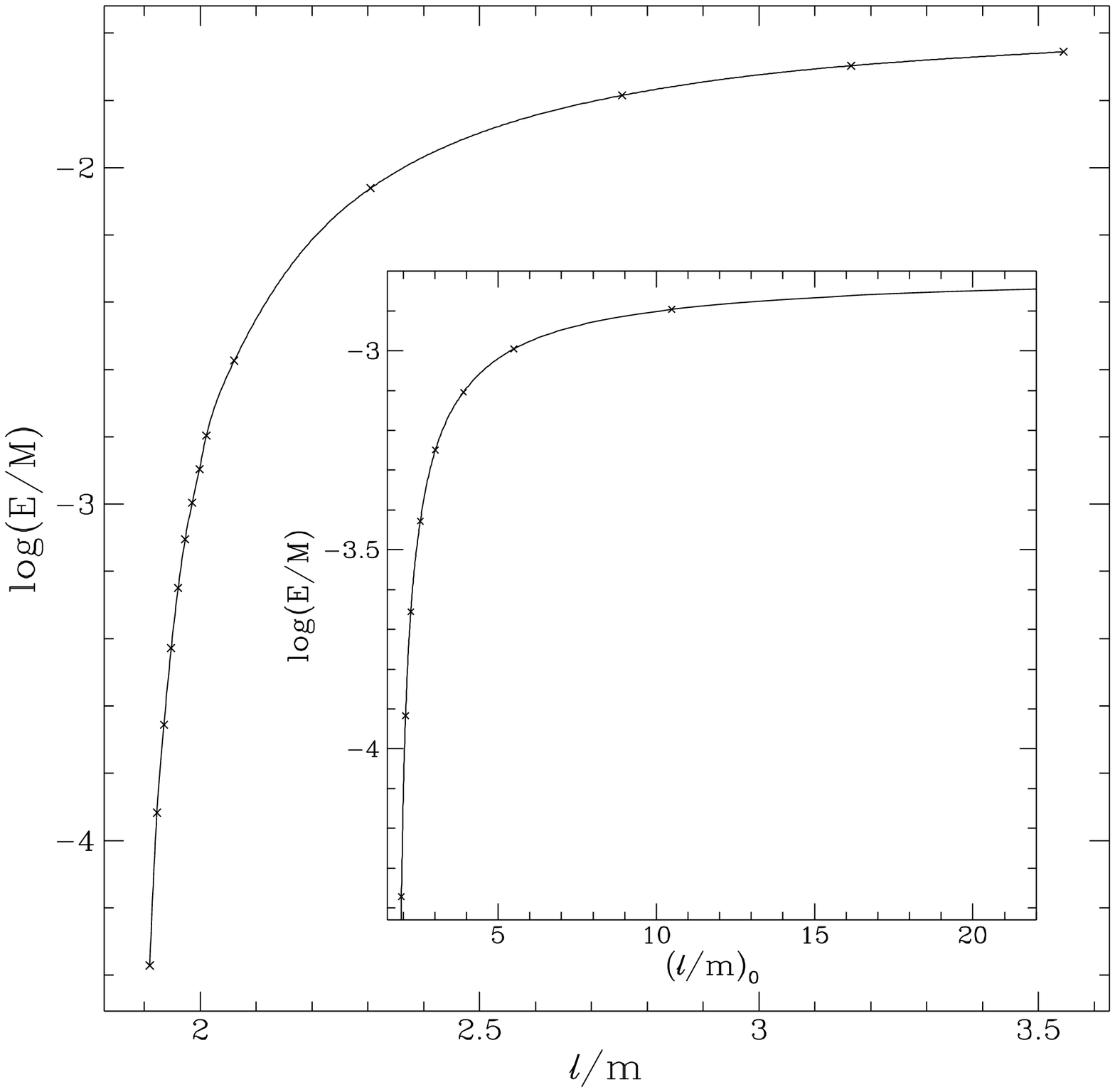}
\vspace{3.4in}
\caption{Radiated energy along the apparent horizon
formation line.  The radiation efficiency (the total
radiated energy as a fraction of the ADM mass)
computed with perturbation theory is plotted
for initial data along the horizon line,
parameterized by the proper separation of the
holes $\ell/m$.  In the inset we show the
radiation efficiency plotted as a function of
the separation of the corresponding time-symmetric initial-data set,
$(\ell/m)_0$, computed using Eq.~(3).}
\label{fig:hleff}
\end{figure}

In Paper 1 we located the horizon-formation line for boosted
black-hole initial data.  For a given separation parameter $\beta$,
we searched for the smallest value of inward linear-momentum for
which an
encompassing horizon surrounded the holes.   The momentum
as a function of proper hole separation for this
horizon line is displayed in Fig.~\ref{fig:hl}.
The horizon cutoff
point mentioned previously, at $\beta=4.17$ and $P=0$, lies on this
line.  For larger values of $\beta$, it is necessary to give
the holes inward momentum in order for an encompassing apparent
horizon to exist.  We note that this horizon line is only an
estimate of where the actual encompassing {\em event}\, horizons
will form.  Along this horizon-formation line, we have computed
the radiated energy for the initial-data sets using the
gauge-invariant perturbation formalism and Zerilli-equation
integration method described above. In Fig.~\ref{fig:hleff}, the
radiated energy is plotted as a function of proper
separation $\ell/m$.
For the points shown, the inward
momentum on each hole ranges
from $P/m=0.0355$ to $P/m=1.738$.  One interesting
feature is that the radiation efficiency appears to
saturate at about $2\%$, substantially below the
maximum radiation efficiency based on area theorem arguments.
This suggests that it may be impossible to obtain high radiation
efficiency for black-hole collisions, even if they
merge with very large momenta.  As a gauge of what constitutes
a large momentum, we estimate below the momenta of two holes
at the moment of horizon formation assuming a parabolic infall
from rest at infinite separation.

At each point on the horizon formation line, we have a value
for the separation $\ell/m$ and momenta $P/m$ of the holes.
Treating the black holes as point particles and using Newtonian
dynamics, we can estimate the separation $(\ell/m)_0$ at which
the holes were at rest:
\begin{equation}
\label{eqn:newt_sep}
	\left({\ell\over m}\right)_0 = {\ell/m\over
		1 - 8 (P/m)^2(\ell/m)}.
\end{equation}
Clearly, if we assume infall from rest, the maximum momenta the
two holes can obtain at the point of horizon
formation is estimated by the point on the horizon-formation
line where the denominator of Eq.~(\ref{eqn:newt_sep}) vanishes,
i.e.\ $(\ell/m)_0\to\infty$.  We find this point to be $P/m = 0.249$
and $\ell/m = 2.01$. At this point, the implied radiation efficiency
is less than 0.16\%.  In the inset of Fig.~\ref{fig:hleff},
we show the radiated energy from points on the horizon-formation
line plotted as a function of the Newtonian estimate
for their proper separation when at rest.
We find striking agreement between our
calculation of total energy radiated and the simulations
of Anninos {\it et al.}\cite{anninos_etal93}.  For cases
where their initial data had an encompassing
horizon, it is clearly correct to compare with the
perturbation analysis of Misner data.  For cases with
greater separation, the time-asymmetric analysis
gives excellent results.  For example, for $\mu_0=2.2$
or $(\ell/m)_0=3.97$, the radiation efficiency
from perturbation theory of the corresponding
horizon-line initial-data set is $E/M=7.9 \times 10^{-4}$,
as compared with $E/M=1.7 \times 10^{-3}$ from the
time-symmetric analysis of $\mu_0=2.2$ Misner data and
$E/M\simeq 5.5 \times 10^{-4}$ from the fully relativistic
simulations.  A post-Newtonian calculation
of the infall trajectory might improve this comparison.

\begin{figure}
\special{hscale=0.45 vscale = 0.45 hoffset = -0.2 voffset = -3.50
psfile=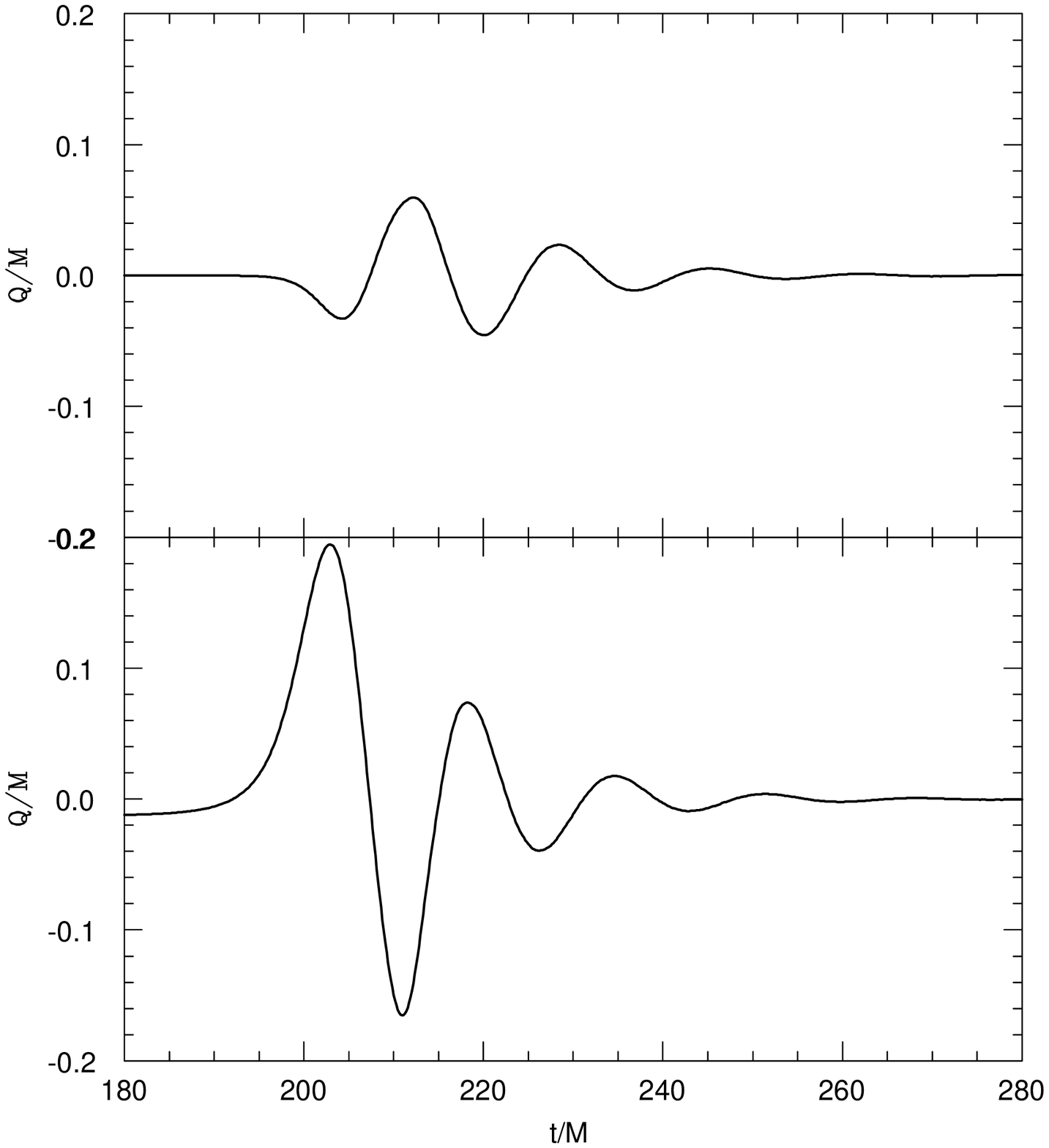}
\vspace{3.5in}
\caption{Waveforms from time-symmetric and
asymmetric two-black-hole initial data.  The top curve
shows the quadrupole wave form from analysis
of a time-symmetric data set with $\beta=4.275$. The
lower curve shows the quadrupole wave form from analysis
of a time-asymmetric initial data set with the same value
of $\beta$ and $P/m=0.1225$.  Both wave forms are
extracted at a radius $r=200M$.}
\label{fig:waveform}
\end{figure}

In Fig.~\ref{fig:waveform} we show a typical wave form
from a boosted head-on collision observed at a radius
of $r=200M$.
The case shown is for a separation
of $\beta=4.275$, $P/m=0.1225$, and $\ell/m=1.948$.
For comparison we show the wave form from time-symmetric data with the
same $\beta$.  Both wave forms are dominated by normal
mode oscillations within about 10M after the black-hole surface
is causally apparent at the extraction radius.   The addition
of ingoing momentum considerably increases the amplitude of the
oscillation and reverses the sign of the wave form.   The
presence of momentum (extrinsic curvature) on the initial slice
causes a significant transient feature in the wave form; this
qualitative effect is not seen by the numerical relativity
simulations of time-symmetric initial data
with $\mu_0=2.2$ because it takes finite time for the momentum
component of the perturbation to propagate out to the
extraction surface. It is possible that this feature will be
present in the evolution of time-symmetric initial data starting
at larger separations.

\section{Discussion}
\label{sec:discussion}
Anninos {\it et al.}\cite{anninos_etal93} have shown that for
collisions resulting from large initial separations there is
excellent agreement between the emitted energy and the well-known
results for a test-particle falling into Schwarzschild
corrected for equal-mass objects, finite infall distance,
and horizon heating.
Price and Pullin\cite{price_pullin94} demonstrated that a
perturbation analysis of time-symmetric initial data could
reproduce the results of the fully relativistic simulation
in the case that the black holes have small initial separation.
Here we have shown that the perturbation analysis can be extended
to larger separations, including the regime in which point particle
analysis is valid, by considering appropriate time-asymmetric
initial-data sets.  Adopting this perspective,
one can accurately predict the total emitted energies
over the entire range of separation, from the close limit
to parabolic trajectories starting at infinity.

\acknowledgments
We would like to thank Richard Price and Jorge Pullin
for pointing out an error in the original version of the paper.
This work was supported by National Science Foundation
grants AST 91-19475 and PHY 90-07834 and
the Grand Challenge grant NSF PHY 93-18152 / ASC 93-18152
(ARPA supplemented).  Computations
were performed at the Cornell Center for Theory and
Simulation in Science and Engineering, which is supported
in part by the National Science Foundation, IBM Corporation,
New York State, and the Cornell Research Institute.
\newpage

%
%
%
%

\newpage

\end{document}